# Standing shock prevents propagation of sparks in supersonic explosive flows

Jens von der Linden [1,2 ✉], Clare Kimblin [3], Ian McKenna[3], Skyler Bagley[4], Hsiao-Chi Li[4], Ryan Houim[4], Christopher S. Kueny [1], Allen Kuhl[1], Dave Grote[1], Mark Converse[1], Caron E. J. Vossen[5], Sönke Stern[5], Corrado Cimarelli [5] & Jason Sears [1]

Volcanic jet flows in explosive eruptions emit radio frequency signatures, indicative of their fluid dynamic and electrostatic conditions. The emissions originate from sparks supported by an electric field built up by the ejected charged volcanic particles. When shock-defined, low-pressure regions confine the sparks, the signatures may be limited to high-frequency content corresponding to the early components of the avalanche-streamer-leader hierarchy. Here, we image sparks and a standing shock together in a transient supersonic jet of micro-diamonds entrained in argon. Fluid dynamic and kinetic simulations of the experiment demonstrate that the observed sparks originate upstream of the standing shock. The sparks are initiated in the rarefaction region, and cut off at the shock, which would limit their radio frequency emissions to a tell-tale high-frequency regime. We show that sparks transmit an impression of the explosive flow, and open the way for novel instrumentation to diagnose currently inaccessible explosive phenomena.

[1] Lawrence Livermore National Laboratory, Livermore, CA, USA. [2] Division E4, Max-Planck-Institut für Plasmaphysik, Garching, Germany. [3] Special Technologies Laboratory, Santa Barbara, CA, USA. [4] University of Florida, Gainesville, FL, USA. [5] Ludwig Maximilian University of Munich, Munich, Germany.
✉email: jens.von.der.linden@ipp.mpg.de

In nature, electrical discharges are frequently observed in widely diverse environments that, beside the common occurrence in thunderclouds[1], include also volcanic plumes[2] and other turbulent particle-laden flows such as dust devils[3], on Earth and other planets. The underlying processes are regulated by the mechanism of induction and separation of electrical charges. Upon electrical discharge, radio frequency (RF) emissions can be recorded, thus providing a means to track the progressive evolution in space and time of the discharge source. Analogous to the detection of thunderclouds and storms, RF detection is now also being used to detect, and inform on the hazards associated with ash-laden volcanic plumes and ash-clouds. In particular the occurrence of electrical discharges at active volcanoes under unrest can be regarded as an indication of the onset of hazardous explosive activity and the production of ash plumes[4–6]. In addition, both observable discharges and RF emissions can reveal the mechanisms that initiate the discharges[7]. The broad RF spectrum associated with lightning discharges results from cascading processes on a hierarchy of time and spatial scales[8,9]. Electric fields accelerate electrons, creating ionization avalanches[10]. A single, or several merging avalanches can collect enough space charge to form a streamer, and such streamers may merge to form a hot self-sustaining plasma channel: a leader. Avalanches and streamers emit very high frequencies (VHF) and leaders emit bright flashes of light together with lower frequencies[1,11,12].

Nature points us to examples where supersonic flows and shocks from explosive events may suppress parts of the hierarchy of the discharge phenomena, such as leaders[13]. In particular, explosive volcanic eruptions produce supersonic flows through the sudden release of overpressured gases contained in the erupting magma, resulting in shock waves. Observation of erupting volcanoes in Alaska[14,15], Iceland[11], and Japan[13] have revealed that in the first few seconds following the onset of an explosive eruption, RF signatures distinct from those produced by leader-forming lightning are recorded in the vicinity (within 10's to 100's of meters) of volcano vents. This early quasi-continuous RF emission is called continual radio frequency (CRF). CRF consists of discrete VHF RF spikes, occurring at rates of tens of thousands per second. Lower frequencies are absent during most of the duration of the CRF although they do occur sporadically, and coincide with prominent visual discharges. These observations suggest that supersonic shock flows may alter the breakdown process hierarchy, so that frequent electrical discharges are occurring with only sporadic leader formation[13]. The hot, opaque plume makes it difficult to determine how the discharges are altered.

Rapid decompression shock tube experiments allow us to explore explosive flows in the laboratory[16–18]. In such experiments, a shock tube ejects a flow of gas and particles into an expansion chamber. Images of non-illuminated decompression reveal bright sparks that are mostly vertical immediately above the nozzle of the shock tube, but bend horizontally at a certain height[19–21]. Reference 22 suggests that the barrel shock structure of a high pressure outflow localizes sparks. Here, we report simultaneous imaging of the Mach disk and coinciding spark discharges, and we provide results of fluid dynamic and kinetic simulations describing the shock flows and breakdown processes. The spatial and temporal scales of the sparks convey an impression of the shock tube flow and kinetic simulations indicate that conditions for discharge are most favorable just upstream of the Mach disk.

## Results

**Optical observations**. To better understand the peculiar localization of spark discharges in the supersonic jet flow, we imaged the evolution of gas-particle mixtures under rapid decompression. In contrast to previous shock tube experiments designed to investigate the role of grain size distribution[20], mass eruption rate[20], and water content[21] in the electrification of volcanic jet analogs, here we focus on the time-space relationship of the discharges and the standing shock, with the objective of informing computer simulations. A series of shock-tube experiments was carried out using mixtures of argon gas and micrometric unimodal diamond powder with nominal average diameters of 5, 50, 250, and 500 μm. Rapid expansion experiments were performed with starting confining pressures of 6.9 and 8.9 MPa depending on the strength of the diaphragms used (Supplementary Table 1). In contrast to previous experiments that used multiple grams of particulate samples to study the dependence of the spark discharges on the standing shock[22], here we use a reduced load of particles to enhance the transparency of the flow and the imaging of the shock barrel evolution over time.

The experimental setup (Fig. 1a) consists of a shock tube (~2.5 cm inner diameter) connected through a nozzle to a large expansion chamber where cameras and antennas diagnose the emerging plume. The images in Fig. 1b–f image a space extending 9.9 cm above the nozzle and show a plume of argon entrained with <100 mg unimodal natural micro diamond powder (~5 μm diameter) acting as charge carriers released through the nozzle into ambient air (2019-03-08 10:15 in Supplementary Table 1 and data appearing in Figs. 1–4 and Table 1, Supplementary Note 1). A time sequence displayed in Fig. 2 shows the fast decompression process with a condensation cloud forming above the nozzle, rising, and then falling, while sparks appear inside the cloud, outlining its sharp upper boundary (Supplementary Movies 1 and 2). Time is referenced to the first pressure increase at the nozzle. In the first 300 μs (blue squares) the condensation appears in the field of view of the camera, rising in height with a sharp top boundary consisting of a flat center and faintly triangular edges, more evident in ensuing images. The condensation boundary then rises above the field of view; but reappears and declines after 1 ms (orange squares). The sparks are localized inside the bounding condensation cloud in the $t = 1.19$ ms and $t = 1.41$ ms frames (red squares, and expanded in Fig. 1b–e). Visible sparks trace the top edge of the condensation boundary including the left edge triangle ($t = 1.19$ ms) and appear vertically through the volume of condensation ($t = 1.41$ ms). Another spark is visible in the $t = 1.33$ ms frame of camera 2 (Supplementary Fig. 1). After 1.5 ms the decline of the condensation slows down, and the boundary (including triangular edges) becomes sharper (black squares, and expanded in Fig. 1f).

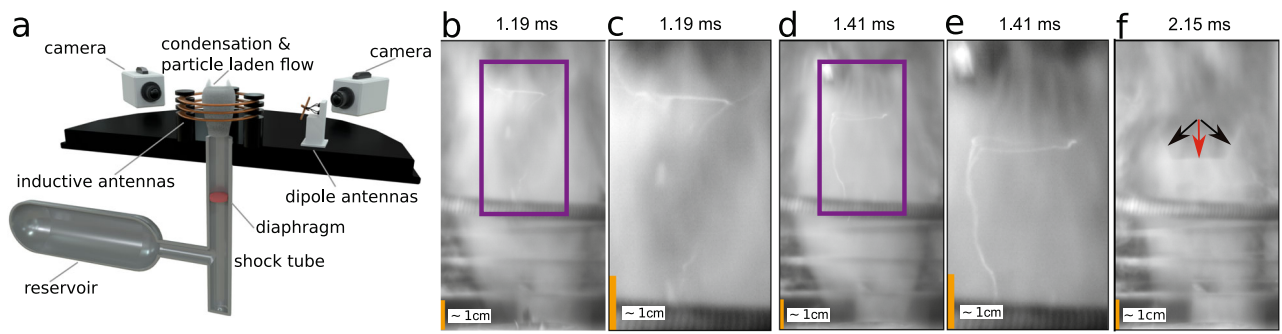

**Fig. 1 Sparks occur below a sharp boundary of the condensation in fast decompression experiments. a** Layout of fast decompression experiment. A diaphragm (red) separates a tube into two sides. The bottom side is filled with varying amounts of particles below an inlet connecting to a gas reservoir, and is pumped to high pressure. The top side of the shock tube connects to an expansion chamber held at atmospheric pressure. The diaphragm bursts at a prescribed pressure, letting the particle-laden gas expand along the tube to the nozzle where a plume and electrical sparks are observed by high-speed cameras, and inductive (ring) and dipole antennas. The gas reservoir delays depletion of the high-pressure gas. **b, d** Images of condensation and sparks in low particle content, fast decompression. **c, e** Enlarged view of the purple-boxed area containing the visible spark. **f** Image of the condensation at later time with sharper flat top boundary (marked by red arrow) and triangular edges (marked by black arrows). Both the condensation and the sparks have a ring-like horizontal upper boundary. Triangular edges are visible in the condensation and the sparks shown in **c**. The time $t$ refers to time since the initial pressure increase at the nozzle.

**Fluid dynamic models.** To relate the boundary to fluid dynamics we perform simulations of the argon gas flow in the full geometry, including reservoir, shock tube, and expansion chamber (simulated pressure and temperature Supplementary Movie 3). Table 1 records the timing of the sequence of events in the experiment in Fig. 1 and the simulation. The argon gas expands through the shock tube as described by fluid dynamic characteristics with the shock, the contact surface between the expanding argon and the low pressure air, and the rarefaction wave propagating through the tube (Fig. 3b). When the gas reaches the nozzle, its pressure is higher than that of the ambient air. When the exit pressure $P_e$ of an expanding gas greatly exceeds the background pressure $P_\infty$ ($P_e/P_\infty > 4$) a singular Mach disk forms[23] (Fig. 3a). In the simulation results this is identifiable as a jump in pressure and temperature (Fig. 4a). The Mach disk boundary in the simulation agrees in shape with the sharp condensation boundary observed in the camera images including the triangular edges. Upstream of the Mach disk the gas expands supersonically, dropping in pressure by 2 orders of magnitude and temperature to below 50 K. While the simulation does not account for phase changes such as heating due to

**Table 1 Timeline of events including Mach disk (MD) rise and decline speed observed in experiment and simulation.**

| Observation | Experiment | Simulation |
| --- | --- | --- |
| Shock passes nozzle | 0 μs | 0 μs |
| CS passes nozzle | - | 120 μs |
| Condensation visible | 150 μs | - |
| MD formation | 300 μs | 360 μs |
| MD rise speed | 170 ± 70 m s$^{-1}$ | 220 ± 10 m s$^{-1}$ |
| MD exits field of view | 440 μs | 540 μs |
| Rarefaction wave passes nozzle | - | 540 μs |
| Reflected rarefaction wave passes nozzle | - | 800 μs |
| MD re-enters field of view | 1.0 ms | 1.26 ms |
| Visible spark 1 | 1.19 ms | - |
| Visible spark 2 | 1.33 ms (Supplementary Fig. 1) | - |
| Visible spark 3 | 1.41 ms | - |
| Initial (0.9–1.5 ms) MD decline speed | − 40 ± 20 m s$^{-1}$ | − 114 ± 1 m s$^{-1}$ |
| MD decline speed after 1.5 ms | − 14.98 ± 0.03 m s$^{-1}$ | − 9.86 ± 0.07 m s$^{-1}$ |

Shock, contact surface (CS), and rarefaction passing refer to the shock tube characteristics exiting the nozzle. The rise and decline speeds are the slopes of linear regressions of the Mach disk height identified in the images and simulations. The measurement of the initial MD decline speed from camera images suffers from the more diffuse boundary in this phase.

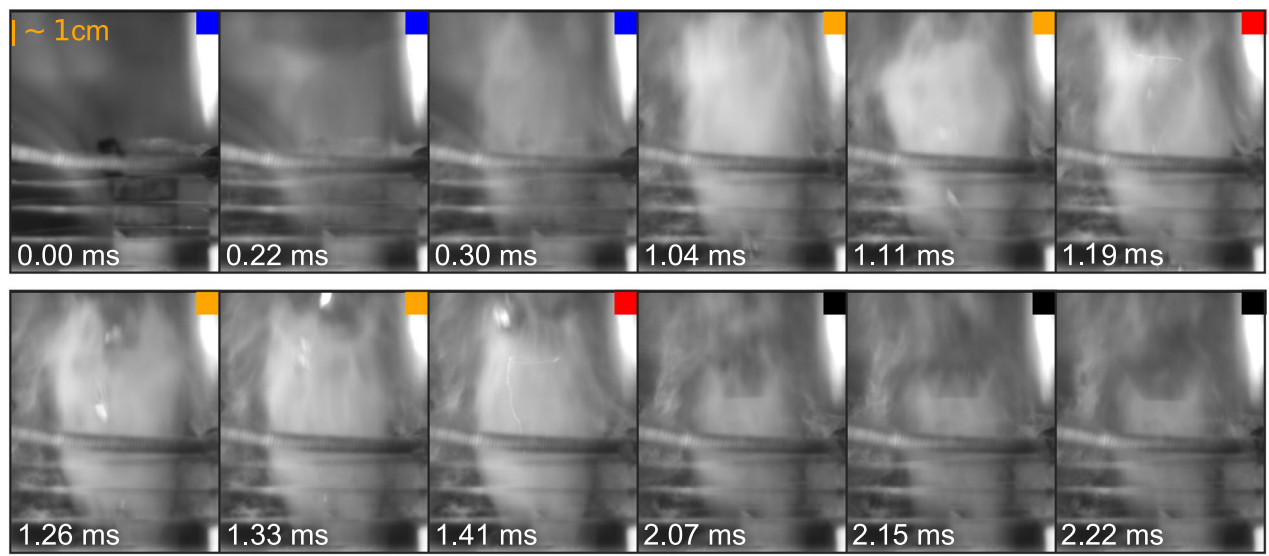

**Fig. 2 Image sequence of sparks and condensation cloud forming a sharp upper boundary, rising, and dropping captured by camera 1.** Camera 1, exposing 20 μs frames at a rate of 13,500 Hz, is triggered by a pressure increase at the nozzle ($t = 0$), and is viewing the nozzle opening and the region above the nozzle including portions of the inductive antennas (10.3 × 8.3 cm shown here, full frame shown in Supplementary Movie 1). Colored squares mark sequential phases of the decompression experiment: rise of condensation and formation of sharp upper boundary (blue), rapid drop of condensation boundary (orange) and electrical sparks (red), slower drop of condensation boundary (black). Sparks are visible in the 1.19 and 1.41 ms frames. Reflective ruptured diaphragm pieces are visible at 1.26, 1.33, and 1.41, and possibly 1.11 ms.

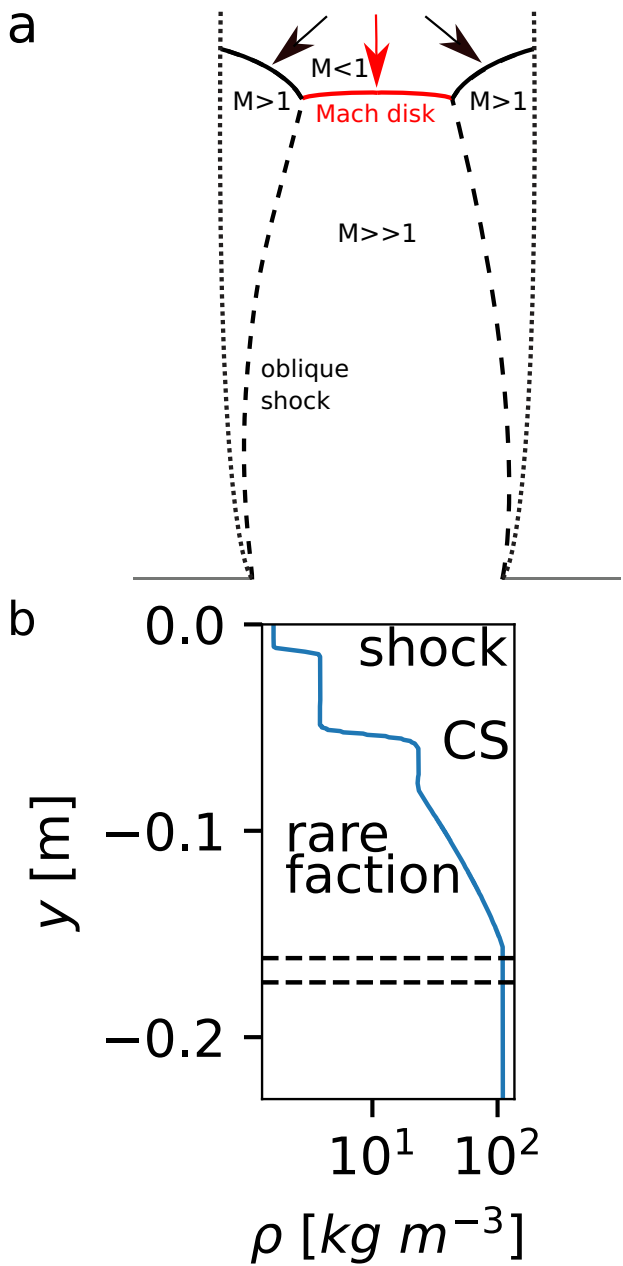

**Fig. 3 Structure of shock flow above and below the nozzle. a** Structure of under-expanded jet above nozzle[23] corresponding to t ~ 1 ms in the experiment and simulation. Oblique shocks launching from the nozzle (long dashed lines) merge to form a flat shock surface, the Mach disk (red line marked with red arrow) separating supersonic rarefaction region from subsonic flow. The reflection of the oblique shocks (solid black line marked with black arrows) form triangular edges with the atmospheric pressure lines (short dashed lines). Discharges are observed in the rarefaction region below the Mach disk enclosed by the oblique shocks. **b** Density along the center axis of the shock tube at t = 0.14 ms before the shock characteristics reach the nozzle at y = 0. The shock and contact surface (CS), and foot of the rarefaction region are moving towards the nozzle. The head of the rarefaction region is propagating towards the bottom of the tube and will reflect. Outflow from the reservoir connection (dashed lines) will later modify the Sod shock tube characteristics.

condensation, this cooling is large enough for the argon to reach temperatures and pressures corresponding to the liquid-vapor phase interface[24]. In this case the presence of particles should enhance inhomogeneous nucleation, making it reasonable to assume that the condensation visible in the images is due to argon condensing[25]. The Mach disk is surrounded by supersonic flows bounded by the shear layer between the expanding gas and the background atmosphere. At the edges of the Mach disk the pressure and temperature isosurfaces are triangular[26] as outlined by the condensation and sparks. Downstream of the Mach disk the flow becomes subsonic and the gas piles up, leading to a pressure jump. The condensed argon vaporizes through the shock wave because of the increment in the temperature and pressure by decreasing the kinetic energy. The contrast between the condensed argon in the cool rarefaction region and its vaporization due to heat of the shock, make its perimeter a tracer of the Mach disk. The simulation reveals the coupling of the Mach disk height to the pressure at the vent (Fig. 4b). As the pressure rapidly increases and the fluid dynamic characteristics of the shock tube flow past the nozzle, the Mach disk forms and rises in both the experiment and simulation by >100 m s$^{-1}$. After 0.5 ms the Mach disk reaches its maximum height in the simulation, rising above the field of view of the camera. During this time, the nozzle pressure reaches its maximum as the reflection of the rarefaction off the bottom of the shock tube reaches the nozzle, after which the nozzle pressure decreases and the Mach disk recedes by <−30 m s$^{-1}$. At about 1.5 ms steadier pressure at the nozzle causes a slow down in the Mach disk decline to ~−10 m s$^{-1}$. This change in pressure at the nozzle is caused by a complex corner flow structure that forms at the right angle entrance of the high-pressure reservoir into the shock tube (Fig. 1a and Supplementary Movie 4). Initially, the outflow from the tube resembles a standard one-dimensional Sod shock tube flow[27]. However, the corner flow creates a standing pressure wave that constricts the outflow from the reservoir and tube below the connection and drives asymmetries in the shock tube flow. In contrast to the outflow from an axisymmetric geometry with equivalent volumes (Supplementary Movie 5) the corner flow creates observable asymmetries in the pressure profiles, which drive unsteady tilting of the Mach disk. Reference [28] differentiates between constant pressure outflows, as "infinite" reservoir flows[29,30], and depleting pressure outflows as "finite" reservoir flows and noted that both can be described with a power law relation[31] for the Mach disk height $h_m$

$$h_m = C d_n (P_n/P_a)^\beta, \quad (1)$$

where $d_n$ is the nozzle diameter, $P_n$ and $P_a$ are the nozzle and ambient pressure, and $C$ and $\beta$ are constants. The observed condensation boundary heights are within the bounds of the power law for an "infinite" reservoir with $C = 0.85 - 0.67$ and $\beta = 0.6$[22], with excursions in height occurring when there is a reversal in sign of the change in pressure (Fig. 4b), possibly indicating hysteresis in the Mach disk height with respect to pressure changes[32]. The agreement between the temporal evolution of the experimentally observed condensation boundary and the simulated Mach disk height are further evidence that the Mach disk regulates the condensation.

**Radio frequency measurements.** Measurements of RF emission suggest that micro-particle charge carriers are necessary to produce sparks during the Mach disk lifetime. With particles present in the flow, both the inductive and dipole antennas respond to intermittent RF pulses throughout the decompression (Fig. 5). The two sparks visible in camera 1 frames of Fig. 1 were imaged at 1.19 ms when RF activity was recorded on the dipole and inductive antennas and at 1.41 ms when RF activity was recorded on the inductive antenna. There are also camera frames with no discernible discharge even though their exposure time overlaps with RF activity. This could be because discharges are obscured by condensation or are occurring inside the tube. There is also considerable RF activity during the inter-frame dead time of the cameras. In contrast, we note that fast decompression of an argon filled shock tube without particles produced only two pulses on the inductive antenna (2019-03-06 13:00 in Supplementary Table 1 and Fig. 5 inset, Supplementary Note 1). These appeared within the first 500 μs and were an order of magnitude weaker in amplitude and probably associated with diaphragm rupture. Following that, no additional pulses were observed on either antenna for the ensuing 10 ms.

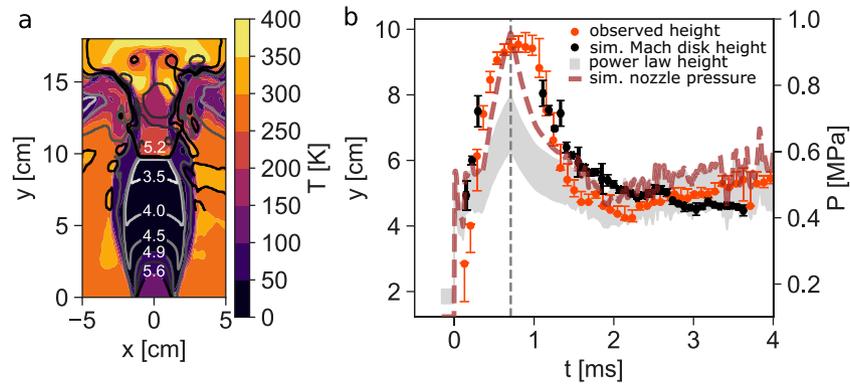

**Fig. 4 Simulation of Mach disk shock agrees with observed evolution of condensation boundary height. a** Simulated temperature in K (filled color contours) and $\log_{10}$ of pressure in Pa (gray contour lines) of fully formed Mach disk above nozzle at $t = 700$ μs after the shock passes the nozzle ($y = 0$ cm). **b** Simulation compared with experiment. Height of Mach disk shock in the simulation (red dots) plotted together with nozzle pressure in the simulation (rust dashes). Error bars in the simulated Mach disk height signify uncertainty due to the uncertainty rating of diaphragm burst pressure and under-resolved features in the corner flow at the reservoir connection. For comparison, the sharp boundary in condensation opacity as identified in the images of the low particle, fast decompression is plotted (black dots). Error bars of experimental Mach disk heights signify high and low points of a linear fit to the sharp condensation boundary in the images. The gray shaded region is the Mach disk power law height calculated with equation (1), the simulated nozzle pressure, $\beta = 0.6$, and the "infinite" reservoir range of C values 0.85−0.67. Dashed gray line marks time of **a**.

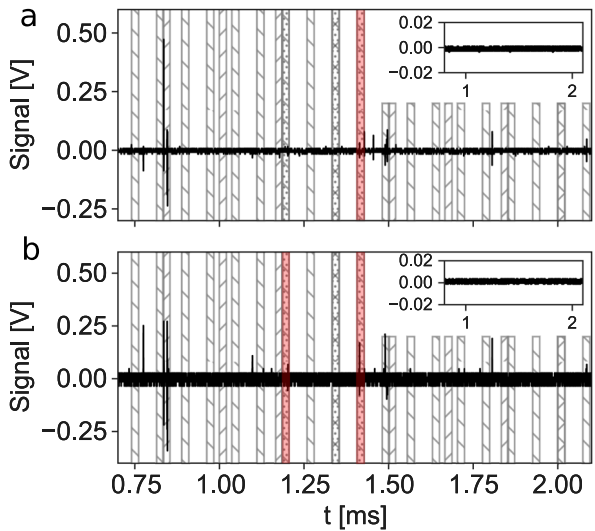

**Fig. 5 RF measurements during the low particle mass, fast decompression.** The exposure times of both high-speed cameras are marked with hatching (camera 1 frames with backward slashes and camera 2 frames with forward slashes). The three frames with visible sparks are marked with dots. The dipole antenna (**a**) records RF coincident with visible sparks at 1.41 ms (red). The inductive antenna (**b**) records RF coincident with visible sparks at 1.19 and 1.41 ms (red). For comparison, the insets show the signal on the respective antennas during a decompression without any particles.

**Electrical breakdown model.** We now turn to analysis of the sparks below the Mach disk. Triboelectric and fracture-charging in the gas and particle outflow can result in a charge imbalance dependent on factors such as particle size distribution, with charge separation then occurring based on inertia[33]. The resulting electric fields accelerate electrons, producing impact ionizations, which generate more electrons. When the electric field is high enough and the gas density low enough to allow for electron acceleration to energies sufficient for impact ionization, this process results in electron avalanches[34]. As charge density increases, the resulting electric field becomes comparable to the external field and a streamer structure arises, with a low electric field in the interior and an enhanced external electric field that drives the advancing ionization front.

The number of electrons in the avalanche $N_e$ is roughly

$$N_e = e^{\int_0^d \alpha(E/n_0) dl}, \qquad (2)$$

where $\alpha$, the Townsend ionization coefficient, is the net number of ionizations generated by a particle per unit length along a path, and the integral is taken over the discharge path $l$ of length $d$. More precisely $\alpha$ is the difference between the Townsend ionization and attachment coefficients; the latter is typically small for conditions of interest here. We have explicitly noted here the strong dependence of $\alpha$ upon the reduced electric field $E/n_0$, where $E$ is electric field and $n_0$ is the local neutral gas number density; ionization is driven by a strong electric field, but will be suppressed if particle mobility is impeded by too high a particle density. Conditions supporting avalanche may be identified via the Raether–Meek criterion[35], which specifies the required number of ionization events along the path. This condition depends upon the gas; for argon it is roughly[36,37]

$$K = \int_0^d \alpha(E/n_0) dl > 10, \qquad (3)$$

where $K$ is the natural logarithm of $N_e$. Reference [22] examined what effect the pressure drop upstream of the Mach disk has on the Paschen condition for Townsend breakdown[38] in experimental volcanic jets similar to those presented here. The Paschen condition is valid for stationary discharges for $pd \lesssim 300$ Pa m, where $p$ is pressure and $d$ is the breakdown distance. Our observed breakdowns are dynamic over distances ranging from 2 to 8 cm in locations with pressure estimated at 0.03–0.05 MPa along the discharge path, making the pressure distance product $pd \sim 500-4000$ Pa m. In this regime the appropriate breakdown condition is the Raether–Meek criterion.

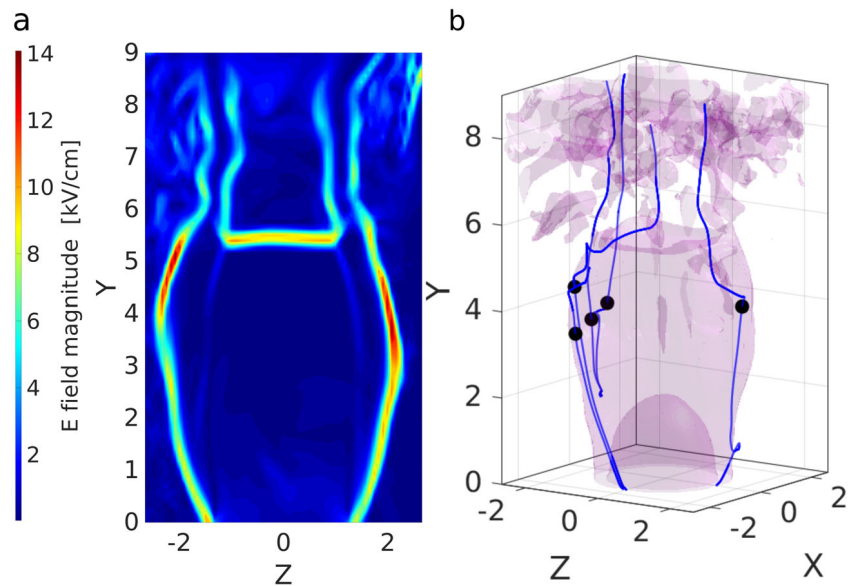

**Fig. 6 Possible discharge paths originating below Mach disk identified by Raether–Meek criterion given in equation (2). a** Model electric field generated as a function of gas velocity field. This model electric field and the fluid simulation data are the inputs to the Raether–Meek criterion. **b** Isosurfaces of gas pressure 0.09 MPa (magenta), with streamlines marking possible breakdown paths. Black dots mark high-$\alpha$ initiation points for streamlines.

In our decompression experiments, the gas density is lowest in the uniform rarefaction region immediately upstream of the Mach disk (Fig. 4a), and this region should therefore be favorable for producing discharges as $E/n_0$ will be high. Sparks in the presence of a Mach disk have been observed to trace out a flat top[20,22] or a circle or semi-circle around the Mach disk, and extend along the edge of the low-density part of the barrel (Fig. 1b–e).

While our fluid dynamic simulations capture gas dynamics well, they do not yet contain a charging model and we are therefore unable to compute electric fields or directly identify breakdown conditions. We can, however, try to surmise the $E/n_0$ required for streamer formation. To get a rough estimate of the required magnitude of $E$, we have computed $\alpha$ for an artificially generated, spatially varying electric field. We might expect charge separation to occur due to varying particle inertia in the direction of fluid flow in regions where this flow is relatively unidirectional. This would support discharges consistent with sparks such as those that are seen along the edge of the barrel in Fig. 1e. For a very rough estimate of the electric field required to produce this spark, we analyze a model field aligned with the fluid flow in this region. Figure 6a shows the magnitude of an electric field defined to be in the direction of gas velocity with magnitude proportional to the velocity gradient, with an arbitrary magnitude scaling factor.

Using this model we have computed the Townsend ionization coefficient $\alpha$ in order to surmise the $E/n_0$ required for streamer formation. Likely breakdown paths are identified by following along the electric field vector through local maxima of $\alpha$. The integral of $\alpha$ along these integration paths gives the total number of ionization lengths, $K$, allowing direct comparison to equation 2. Figure 6b shows the computed breakdown paths with largest $K$ value for the time corresponding to the spark in Fig. 1e ($t = 1.41$ ms). The location of the Mach disk just above 5 cm is marked by the pressure (magenta) isosurface. For this electric field, the maximum $K$ was 23 ionization lengths, with dozens of streamlines having $K > 10$, the rough argon threshold.

Since our simple model produces breakdown paths strictly aligned with the velocity field, it cannot predict structures such as the illuminated ring or flat-top seen along the Mach disk in the experiment (Fig. 1). Recent computational work[39,40] has investigated the effect of density discontinuities on streamers, demonstrating their redirection parallel to the discontinuity. This analysis is relevant to our Mach disk geometry and the ring-like horizontal discharges seen along the Mach disk boundary in Fig. 1c, e. Further analysis of our case awaits implementation of a particle charging model in our simulations.

## Discussion

We observe coincidence between sparks and a sharp boundary in the condensation of a fast decompressing gas loaded with a small quantity of solid particles. The observed sparks result from electrical breakdown and emit RF signals measurable on antennas. Based on the spatial coincidence of rarefaction condensation and electrical sparks, and, based on an analysis of streamer formation criterion upstream and downstream of the Mach disk, we argue that the space occupied by the first shock cell (the distance between the Mach disk and the nozzle exit) regulates the space in which electrical breakdown occurs in the presence of particles.

Explosive granular flows involve multi-phase flows coupled by drag. Solving these coupled multi-phase equations is complicated by non-conserved nozzeling and $pdV$ work terms[41], and the effect of particle pressure[42,43]. Rapid decompression presents a validation challenge for granular compressible fluid dynamics models, which must represent the transition from the dense granular regime in the tube to the dilute granular regime in the expansion chamber. Imaging of the condensation, before the particles reach the plume, and sparks, after the particle plume obscures the condensation, may provide measurements of the Mach disk shape and height for comparison with models.

Explosive volcanic eruptions are an example of granular flows that could regulate breakdown processes. Experiments[44] and simulations[29] have shown that the pressures released in a volcanic eruption could result in under-expanded flows forming standing shock waves with Mach disks[45]; however, depending on their concentration and Stokes number, particles may perturb or inhibit the Mach disks[30]. If Mach disks do form in volcanic outflows, they may regulate electrical discharges between charged ash particles. The continual radio frequency (CRF) signature[13], distinct due to the lack of low radio frequency emission, which is observed with coincident volcanic lightning[46] appearing near the volcanic vent, may be regulated by the shock flow. If the sources of near-vent CRF emission are indeed regulated by standing shock waves, then distributed antennas could triangulate their locations, tracking the evolution of the regulating standing shock and providing insight into the pressure and particle content of the explosive flow. Our fast decompression experiments and simulations permit observation and analysis of explosive events producing RF at their onset, and may lead to insights into conditions, which favor streamer/avalanche over leader-forming lightning.

## Methods

**Fast decompression experiments.** The fast decompression experiments at the Special Technologies Laboratory (STL) described here are designed and operated as described in refs. [18,19] with two notable exceptions. A 1 l reservoir volume is added to the high-pressure side to ensure the transition to the "infinite" reservoir Mach disk regime. Moreover, for the low particle content shot described here, natural 5 μm diamond powder (Lands Superabrasives, LSNPM 3–7, ~5 μm) was deposited through a previous decompression along the walls of the shock tube and surfaces of the expansion chamber. By comparing the observed particles to shots with mg sample masses in the sample holder we determine the particle mass in the shot described here to be no larger than 100 mg. Argon gas is used to pump the high-pressure side of the shock tube. When the pressure exceeds the pressure at which the diaphragm (Oseco, 6.9 MPa, 2.54 cm diameter STD rupture disc) is rated to burst, the gas and entrained particles are rapidly ejected out of the transparent plastic vent above the baseplate into ambient air in the expansion chamber (with internal diameter 39 cm and height of 137 cm). The compressible fluid dynamics is diagnosed with a pressure transducer at the reservoir, and using camera images of the condensation plume. Camera 1 and 2 are Photron SA4 and SA3 high-speed cameras, respectively. For the shot described here camera 1 imaged a field of view of 11.1 × 11.2 cm, 9.9 cm above the nozzle and camera 2 imaged a field of view of 5.9 × 7.4 cm, 6.4 cm above the nozzle. The electrical activity is diagnosed using camera imaging and self-built inductive (ring) and dipole antennas.

In addition to the shot discussed in this paper, sparks were observed in the condensation during four illuminated argon rapid decompression experiments (6.9 MPa) with only diamond particles, and in one shot with residual 50 μm diamond and 150 μm graphite (15 g) before graphite entrainment was observable. In an illuminated shot with a considerable quantity of 5 μm diamond present (30 g) sparks framed by the condensation were not observed. This may be due to early entrainment of the particles creating an opaque and reflective cloud, which obscured observation of Mach disk framed sparks. In contrast, in two illuminated shots (6.9 MPa) of pure argon, only the sharp condensation boundary was observed without any sparks. With the illuminated blank shot for which RF was recorded, there was no RF activity between 0.3 and 13 ms. In a third pure argon shot with lights off, no sparks were observed at all, and there was no RF activity between 0.28 and 9.7 ms. The sharp boundary in the condensation without sparks was also observed in three gas shots without particles at the Ludwig Maximilian University of Munich (LMU) rapid decompression facility described in ref. [19] with a burst pressure of 8.9 MPa. Notably, the LMU shock tube does not have a reservoir volume, resulting in more transient decompression.

**Image analysis.** A Hough transform[47] of the image was taken to identify and fit the edge in the image intensity caused by the sharp condensation boundary. The endpoints of the most probable edge were fit with a linear equation. The error was taken to be the difference between the heights at the ends of the fit line. To avoid confusion with intensity edges from stationary components such as the antennas, the image frame was cut to a 200 × 232 pixel frame of the central area immediately above the inductive antennas. The known distance between the three inductive antenna rings and a pixel count between the rings determined the cm per pixel resolution of the camera frames.

**Fluid dynamic simulations.** Only the shock tube with the reservoir was simulated here. The simulations of the argon flow were conducted by solving compressible Navier-Stokes equations with the HyBurn code based on the AMRex framework[48]. HyBurn uses a high-order Godunov algorithm with the HLLCM approximate Riemann solver[49] and the seventh-order WENO method[50] to reconstruct the primitive variables (pressure, temperature, and velocity). The solution was marched in time using third-order strong-stability preserving Runge-Kutta[51]. HyBurn implements the complex geometry of the gas reservoir, shock tube, and expansion chamber using immersed boundary methods[52]. First, the geometry was drawn using the SolidWorks computer-aided design package and exported to a stereolithography file. The complex geometry was input into HyBurn by generating a signed-distance function[53] based on the intersection points of triangles in the stereolithography file with the computational mesh. The boundary conditions on the embedded geometry are enforced using the method of images[52]. The uncertainty in the simulation was determined by varying the burst pressure of the diaphragm within the −3% /+6% rating of the manufacturer and by conducting a convergence study. The convergence study revealed that not all features of the flow at the 90° reservoir connection to the shock tube can be resolved; however, the Mach disk evolution is relatively insensitive to the resulting variation in pressure profile due to Mach disk hysteresis[32].

**Kinetic simulations.** The Townsend ionization coefficient $\alpha$ is calculated with the Boltzmann solver BOLSIG+[54]. To calculate $\alpha$, BOLSIG+ needs density, temperature, electric field, and molecular cross-sections. The spatial distributions of density $n_0$ and temperature are obtained from the Hyburn fluid dynamics simulation output. The reduced electric field $E/n_0$ is modeled as described in the results section. Among neglected effects is the influence of the metallic burst disc fragments pointed out in Fig. 2. A future campaign with non-fragmenting burst discs could determine if the model needs to be adjusted. The argon reaction cross sections are obtained from the LXCat[55] Morgan database (retrieved on May 21, 2020). The background ionization to neutral ratio is approximated from the Saha equation as $\sim 10^{-28}$; the results are very insensitive to this number. BOLSIG+ produces as output the reduced ionization coefficient $\alpha/n_0$, which is very strongly dependent on $E/n_0$. The procedure for finding potential breakdown paths begins with tabulating all local maxima of $\alpha$ to use as initiation points for discharges. Paths are then found by computing forward and backward streamlines along the electric field through these points, terminating when $\alpha$ falls below some chosen threshold. Finally, we integrate $\alpha$ along each path to obtain the total number of ionization lengths $K$, allowing direct comparison to equation (2). This model provides a rough estimate of the discharge conditions in the bulk flow.

## Data availability

The data measured in the experiments and produced by the simulations described here is available in the Zenodo repository: the data from the experiments at Special Technologies Laboratory (https://doi.org/10.5281/zenodo.4245225)[56], the data from the experiments at Ludwig Maximilian University of Munich (https://doi.org/10.5281/zenodo.4127362)[57], the outputs of the fluid dynamics simulations (https://doi.org/10.5281/zenodo.4127362)[58], and the outputs from the Boltzmann solver (https://doi.org/10.5281/zenodo.4128164)[59].

## Code availability

The fluid dynamics code HyBurn (commit ec13c04) is available from R.H. upon request. The AMRex framework (commit e915f9e), which HyBurn utilizes is available from the AMRex developers under an open source license (https://github.com/AMReX-Codes/amrex). The Boltzmann solver Bolsig+ (version 12/2019) is available from Gerjan Hagelaar of the LAPLACE laboratory in Toulouse, France (http://www.bolsig.laplace.univ-tlse.fr). The analysis codes used in this study are available from the corresponding author upon request.

## Acknowledgements
We acknowledge the support of staff: Matt Staska, Mary O'Neill, Jonathan Madajian, Ben Valencia, Roy Abbott, and Rick Allison at Special Technologies Laboratory and Markus Sieber at Ludwig Maximilian University of Munich. We acknowledge Taralyn von der Linden for creating the 3D visualization of the experiment layout in Fig. 1a. This work was performed in part under the auspices of the U.S. Department of Energy by Lawrence Livermore National Laboratory under contract DE-AC52-07NA27344, and Mission Support and Test Services, LLC, under Contract No. DE-NA0003624 with support from the Site-Directed Research and Development program, DOE/NV/03624−−0956, and in part by the European Plate Observing Systems Transnational Access program of the European Community HORIZON 2020 research and innovation program under grant N 676564. C.C. acknowledges the support from the DFG grant CI 25/2-1 and from the European Community HORIZON 2020 research and innovation program under the Marie Sklodowska Curie grant nr. 705619. J.v.d.L. acknowledges support from the Alexander von Humboldt foundation. LLNL-JRNL-811941. This document was prepared as an account of work sponsored by an agency of the United States government. Neither the United States government nor Lawrence Livermore National Security, LLC, nor any of their employees makes any warranty, expressed or implied, or assumes any legal liability or responsibility for the accuracy, completeness, or usefulness of any information, apparatus, product, or process disclosed, or represents that its use would not infringe privately owned rights. Reference herein to any specific commercial product, process, or service by trade name, trademark, manufacturer, or otherwise does not necessarily constitute or imply its endorsement, recommendation, or favoring by the United States government or Lawrence Livermore National Security, LLC. The views and opinions of authors expressed herein do not necessarily state or reflect those of the United States government or Lawrence Livermore National Security, LLC, and shall not be used for advertising or product endorsement purposes.

## Author contributions
J.v.d.L. and J.S. conceived the decompression shots at Special Technologies Laboratory and Ludwig Maximilian University to identify Mach disk tracers. C.K. and I.M. designed the decompression experiment at Special Technologies Laboratory. C.K., I.M., J.v.d.L., and J.S. operated and analyzed the Special Technologies Laboratory experiments. R.H. developed the HyBurn compressible fluid dynamics code. S.B. adapted the HyBurn compressible fluid dynamics code to the experiment geometry. H.C.L. implemented the immersed boundaries in HyBurn needed to capture the full geometry. D.G. and A.K. validated the HyBurn code for the explosive flow conditions. J.v.d.L. ran HyBurn on high performance computing resources and analyzed the outputs. C.S.K. analyzed the simulation outputs with the Bolsig+ Boltzmann solver to determine ionization rates. M.C. evaluated particle charging mechanisms. J.v.d.L., C.C., and J.S. designed the experiments at Ludwig Maximilian University of Munich. C.E.J.V., S.S., J.v.d.L., J.S., and C.C. operated and analyzed the Ludwig Maximilian University experiments. J.v.d.L. wrote the manuscript and all authors provided input.

## Competing interests
The authors declare no competing interests.

## Additional information

**Supplementary information** The online version contains supplementary material available at https://doi.org/10.1038/s43247-021-00263-y.

**Correspondence** and requests for materials should be addressed to Jens von der Linden.

**Peer review information** *Communications Earth & Environment* thanks Juan José Peña Fernández and the other, anonymous, reviewer(s) for their contribution to the peer review of this work. Primary Handling Editor: Joe Aslin. Peer reviewer reports are available.

**Reprints and permission information** is available at http://www.nature.com/reprints

**Publisher's note** Springer Nature remains neutral with regard to jurisdictional claims in published maps and institutional affiliations.